\documentclass[12pt, preprint]{aastex}
\usepackage{bbm}
\usepackage{natbib}
\usepackage{rotating}
\usepackage{amsmath}

\begin{document}

\title{Extragalactic Point Source Search in WMAP 61 and 94~GHz Data}

\author{X. Chen \& E. L. Wright}
\affil{Physics and Astronomy Department, University of California,
	Los Angeles, CA 90095-1547}

\begin{abstract}

We report the results of an extragalactic point source search using the 61 and 94 GHz (V- and W-band) temperature maps from the Wilkinson Microwave Anisotropy Probe (WMAP).
Applying a method that cancels the ``noise'' due to the CMB anisotropy signal, we find in the $|b| > 10\degr$ region 31 sources in the first-year maps and 64 sources in the three-year co-added maps, at a $5\sigma$ level.
The 1$\sigma$ position uncertainties are 1.6$^\prime$ and 1.4$^\prime$ each.
The increased detections and improved positional accuracy are expected from the higher signal-to-noise ratio of WMAP three-year data. All sources detected in the first-year maps are repeatedly detected in the three-year maps, which is a strong proof of the consistency and reliability of this method.
Among all the detections, 21 are new, i.e. not in the WMAP three-year point source catalog.
We associate all but two of them with known objects.
The two unidentified sources are likely to be variable or extended as observations through VLA\footnote{The Very Large Array. See~\url{http://www.vla.nrao.edu/}.}, CARMA\footnote{The Combined Array for Research in Millimeter-wave Astronomy. See~\url{http://www.mmarray.org/}.} and ATCA\footnote{The Australia Telescope Compact Array. See~\url{http://www.narrabri.atnf.csiro.au/}.} all show non-detection at the nominal locations.
We derive the source count distribution at WMAP V-band by combining our verified detections with sources from the WMAP three-year catalog.
Assuming the effect of source clustering is negligible, the contribution to the power spectrum from faint sources below 0.75 Jy is estimated to be $(2.4\pm0.8) \times 10^{-3} \mu$K$^2$~sr for V-band, which implies a source correction amplitude $A = 0.012\pm0.004~\mu$K$^2$~sr.

\end{abstract}

\keywords{cosmic microwave background --- cosmology: observations --- methods: data analysis}

\section{Introduction}

The Wilkinson Microwave Anisotropy Probe was launched in 2001 for making precise measurements of the CMB anisotropy (\citealt{2003ApJ...583....1B}).
Since on small scales the CMB signal in the WMAP full-sky maps is primarily contaminated by microwave emission from extragalactic point sources, it is important to detect and mask out point-source-contaminated pixels\footnote{The Hierarchical Equal Area isoLatitude Pixelization (HEALPix) of the sphere is used to define WMAP map pixels on the sky in Galactic coordinates. A resolution level r=9 is chosen, which corresponds to 3,145,728 pixels with a pixel resolution of $0.115\degr$~(\citealt{2003ApJS..148....1B}). }.
Accordingly, two extragalactic point source searches were made in WMAP first-year maps and three-year co-added maps\footnote{Data products are available on-line through the LAMBDA website: \url{http://lambda.gsfc.nasa.gov/product/map}.} by the WMAP science team~(\citealt{2003ApJS..148...97B}, \citealt{2007ApJS..170..288H}).
The temperature maps were first weighed by $N_{obs}^{1/2}$ and then filtered by $b_{l}/(b_{l}^{2}C_{l}^{CMB}+C_{l}^{noise})$ in harmonic space, where $N_{obs}$ is the number of observations, $b_{l}$ is the WMAP beam transfer function, $C_{l}^{CMB}$ is the CMB angular power spectrum and $C_{l}^{noise}$ is the noise power.
Peaks that are greater than $5\sigma$ in any band of the filtered maps were interpreted as source detections.
This procedure yielded a catalog of 208 point sources in the first-year data and an enlarged catalog of 323 point sources in the three-year data, with a position uncertainty of 4$^\prime$ for both searches.
A point source mask of nearly 700 objects was then constructed to include all the 323 WMAP directly detected sources, along with the sources from \citet{1994A&AS..105..211S}, sources with 22 GHz fluxes $\ge$ 0.5 Jy from \citet{2000PASJ...52..997H}, flat-spectrum objects from \citet{2001A&A...368..431T}, and sources from the X-ray/radio blazar survey of \citet{1998AJ....115.1253P} and \citet{2001MNRAS.323..757L}.
Each source was masked to a radius of 0.6\degr.
On top of that, the power spectra of the unresolved sources were modeled into 
\begin{equation}
C_l^{(i, i^\prime)}~ =~A~(\frac {\nu_i}{\nu_Q})^{\beta}~(\frac {\nu_{i^\prime}}{\nu_Q})^{\beta}~\omega_l^{(i, i^\prime)},
\end{equation}
with $A = 0.014\pm0.003~\mu$K$^2$~sr, $\beta = -2.0$, $\nu_Q = 40.7$~GHz.
Here~$\omega_{l}$ is the window function that gives the combined smoothing effects from the beam and the finite sky map pixel size, and $(i, i^\prime)$ denotes a pair of DA indices~(\citealt{2003ApJS..148..135H, 2007ApJS..170..288H}).
This model was subtracted from the WMAP derived power spectra preceding any cosmology analysis.

In this paper, we present a new extragalactic point source search in the WMAP V- and W-band data, using a method that has no CMB dependence.
The main purpose of this work is to find more point sources in these two bands by lowering the noise level as only 76.2$\%$ and 37.5$\%$ of the WMAP three-year cataloged sources are identified with~$> 2\sigma$~confidence in V- and W-band, respectively.
Secondly, we are interested in seeing sources that might peak at these high frequency bands, for instance, the so-called Gigahertz-Peaked Spectrum (GPS) sources which are believed to be at the very early state of the evolution of powerful radio sources~(\citealt{1998PASP..110..493O}).
A third motivation arises because in a recent re-analysis of WMAP temperature maps by~\citet{2007ApJ...656..641E}, an unexpected systematic discrepancy was found between the V- and W-band power spectra on small angular scales~(few percent at $\ell \ga 300$).
This was partly explained by \citet{2006ApJ...651L..81H} to be an over-subtraction of unresolved point sources, and the significance of this difference was reduced from 3$\sigma$ to 2$\sigma$ by applying a smaller source correction~$A = 0.011\pm0.001~\mu$K$^2$~sr.
This amplitude was later revised to $A = 0.012\pm0.005~\mu$K$^2$~sr for $\ell < 500$ and $A = 0.015\pm0.005~\mu$K$^2$~sr for $\ell >500$~in~\citet{2007arXiv0710.1873H}.
Therefore, we also aim at getting a better constraint on the unresolved source contamination by using direct measurements.

\section{Methodology}

\begin{figure}
\epsscale{0.65}
\plotone{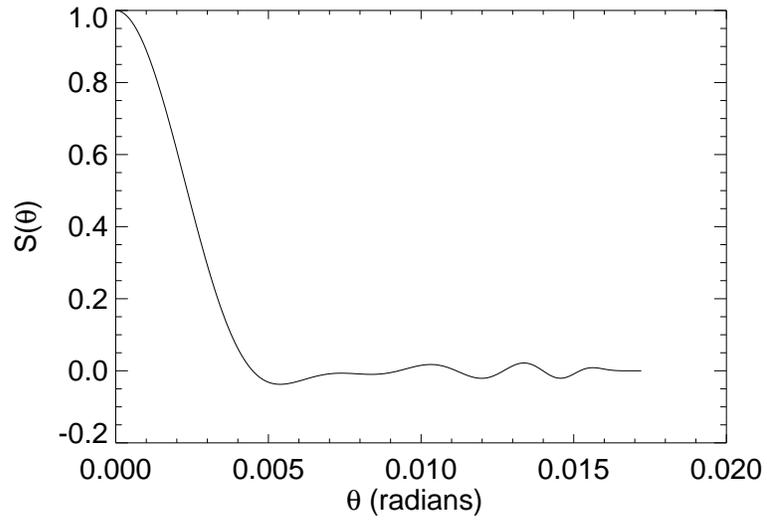}
\caption{Smoothing function S($\theta$), used to degrade the W-band map to V-band resolution.
The cutoff radius $\theta_{R_{c}}$~is 1\degr ($\approx$ 0.017 radians). 
\label{S}}
\end{figure}

We start from the V- and W-band WMAP full sky temperature maps (the CMB dipole has been removed).
Since WMAP observes the sky convolved with different beam patterns at different bands~(\citealt{2003ApJS..148...39P}), we first smooth the high resolution W-band map to match V-band resolution.
A cutoff radius $\theta_{R_{c}}$ of 1\degr~is chosen for the solid angle integrations.
A smoothing function $S(\theta$) is defined as a polynomial of an infinitely smooth function with compact support,
\begin{equation}
 S(\theta)~=~\sum_n~a_nS_n~(\theta), 
\end{equation} where
\begin{equation}
S_n~(\theta) = 
\begin{cases}
\exp(-~\frac {n \theta^2}{\theta_{Rc}^{2}~-~\theta^2}) & \text{$\theta~<~\theta_{Rc}$} \\
0 & \text{$\theta~>~\theta_{Rc}$}.
\end{cases}
\end{equation}
The coefficients $a_{n}$ are found by minimizing the difference between the Legendre transform of the smoothing function
\begin{equation}
S_l~=~2 \pi \int S(\theta) P_l(\cos \theta) d (\cos \theta)
\end{equation}
and the ratios of WMAP V- and W-band beam transfer functions
\begin{equation}
 S_l~=~b_l^V / b_l^W.
\end{equation}
We start with the 1$^{st}$ power of the polynomial and truncate it after the
10$^{th}$ power, so only the coefficients $a_1$ to $a_{10}$ are non-zero,
as this is sufficient to give a good fit of $b_l^V / b_l^W$.
The profile of $S(\theta)$~is given in~Figure \ref{S}.
The W-band map is degraded to V-band resolution by convolving with this smoothing function.

\begin{figure}
\epsscale{0.65}
\plotone{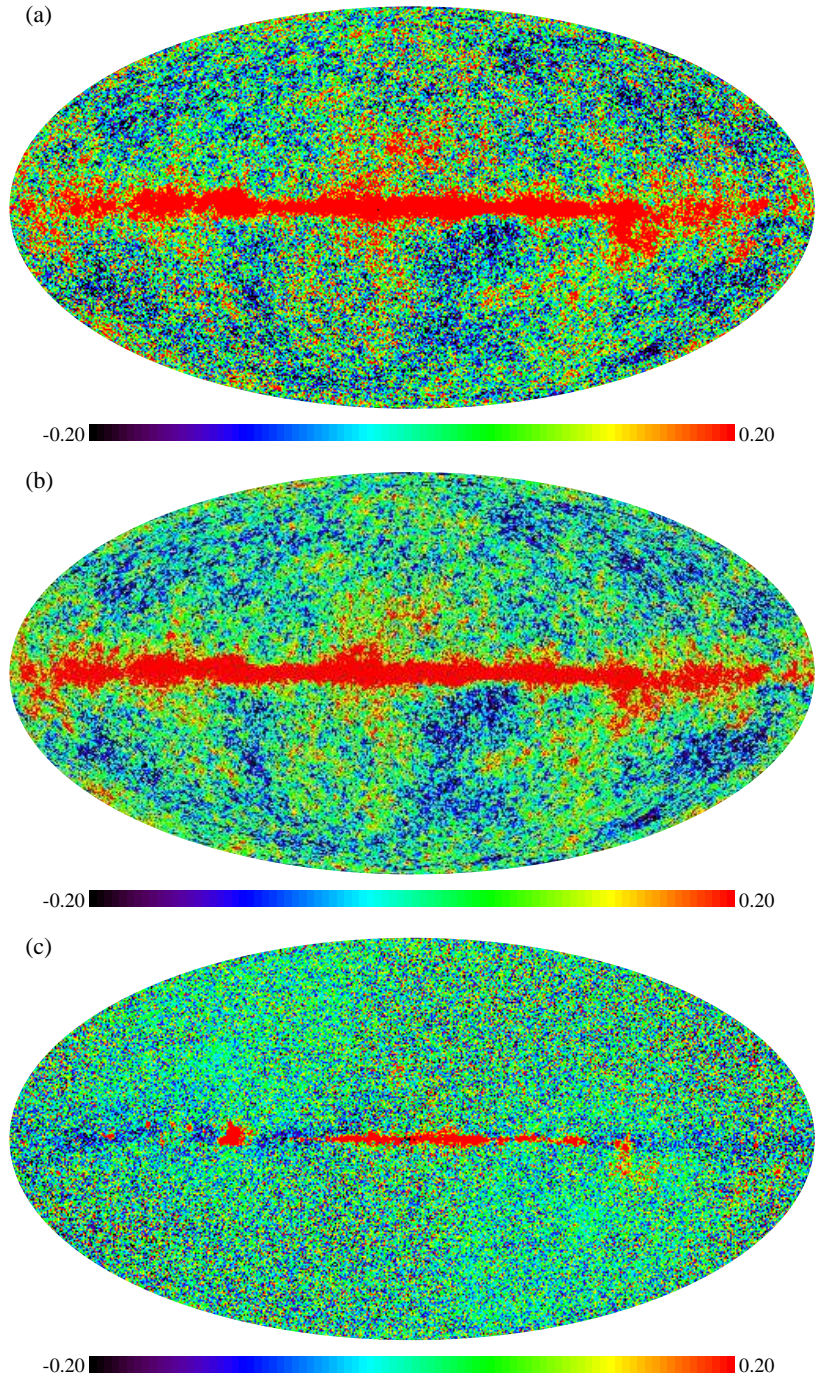}
\caption{(a) The V-band sky map. (b) The W-band sky map smoothed to V-band resolution. (c) The residual sky map after subtracting (b) from (a). Regions around the ecliptic poles and latitude $\pm$ 45$\degr$ are heavily observed by WMAP and thus appear less noisy. The maps are shown in the Mollweide projection in units of CMB thermodynamic temperatures (mK).
\label{mapsub}}
\end{figure}

We then subtract the smoothed W-band map from the V-band map.
Since the CMB anisotropy is from gravitational perturbations and Thomson scattering of blackbody radiation, all anisotropy-induced ``noise'' is removed at this step.
One assumption we made here is that the point sources we are looking for do not have a blackbody spectrum.
This is a reasonable assumption as the sources observed at WMAP frequencies are found to be predominately flat spectrum, i.e. $F_{\nu} \propto \nu^{0}$ (\citealt{2003ApJS..148...97B}).
The V-band map, smoothed W-band map and the residual map are shown in Figure \ref{mapsub}.
Regions around the ecliptic poles are seen to be less noisy compared to other parts in the sky since these areas were heavily observed by WMAP. And because the Galactic emission in V-band is free-free dominated and in W-band is dust dominated, the residual map also reflects the ratio of free-free emission to dust emission, i.e., red represents strong free-free emission areas and blue represents strong dust emission areas. 

\begin{figure}
\epsscale{0.65}
\plotone{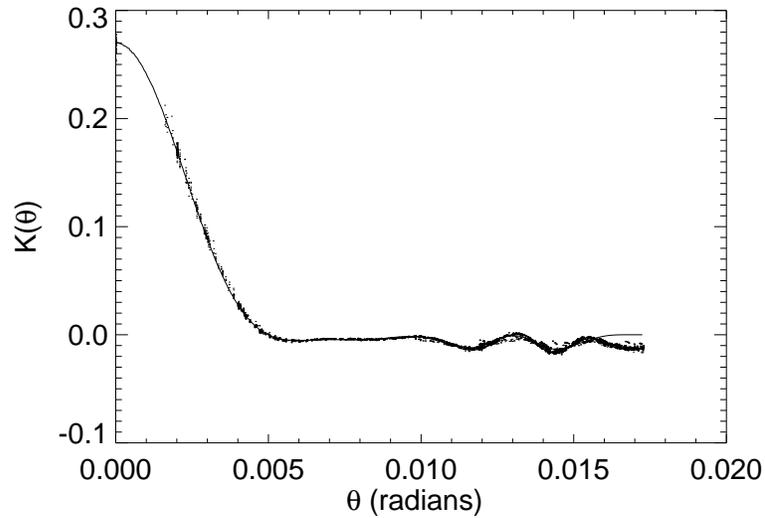}
\caption{The filter we used to minimize the noise and recover the point source flux distribution in the V-W residual map. Dots are superimposition of generated data at 32 random sky locations;~Solid line is our fit. The scale of K($\theta$) is determined by the normalization of the V-band beam profile as indicated by Equation (6) and Equation (12).
\label{kernel}}
\end{figure}

If there is a point source with flux $f$ in the residual map and assuming the contribution from the overlap of other point sources is negligible, the resulting intensity at a close-by pixel~$i$~can be fit into
\begin{equation}
V_i ~-~ W_i^{SM} ~=~ f b^V_i ~+~ c_i.
\end{equation} 
Here~$V_{i}$ and $W_{i}^{SM}$ are pixel temperatures in the V-band and smoothed W-band maps;~$b^{V}_{i}$ is short for $b^{V}(\theta_{i})$, where $b^{V}(\theta)$ is the V-band beam profile normalized to 1 at $\theta = 0$ and $\theta_{i}$ is the angular distance from pixel $i$ to the considered source;~$c_i$ is the local baseline.
The $\chi^2$ of the above fit is
\begin{equation}
  \chi^2 ~=~ \sum_{ij} \epsilon_i ( N^{-1}_{ij} ) \epsilon_j,
\end{equation} 
with
\begin{equation}
  \epsilon_i ~=~  (V_i - W_i^{SM}) ~-~ (f b^V_i + c_i)
\end{equation} 
and
\begin{equation}
 N_{ij} ~=~\delta_{ij} \sigma_V^2 ~+~ \sum_k S_{ik} S_{jk} \sigma_W^2.
\end{equation} 
Here~$\sigma_V$ is the noise at pixel $i$ in the V-band map and $\sigma_W$ is the noise at pixel $k$ in the pre-smoothed W-band map;~$S$ is the previously-defined smoothing function.
The summation over
$k$ is a sum over all the pixels within $\theta_{R_{c}}$ of pixel $i$ and $j$ as the W-band noise is correlated after the smoothing process.
A least $\chi^2$ fit then gives
\begin{equation}
\sum_{i j} N^{-1}_{i j} \left(
\begin{array}{c}
b_i \\
1
\end{array}     \right) \left(
\begin{array}{cc}
b_j & 1 \end{array} \right)  \left(
\begin{array}{c}
f \\
c
\end{array}     \right) = 
\sum_{i j} N^{-1}_{i j} (V_j - W_j^{SM}) 	\left(
\begin{array}{c}
b_i \\
1
\end{array}     \right).  
\end{equation}
Defining
\begin{equation}
M  =  \left(
\begin{array}{cc}
M_{00} & M_{01} \\
M_{10} & M_{11}
\end{array}     \right) = \sum_{i j} N^{-1}_{i j} \left(
\begin{array}{cc}
b_i b_j & b_i\\
b_j & 1
\end{array}     \right), 
\end{equation}
the intensity of the source can be written as a weighted sum of the intensities in the surrounding pixels, 
\begin{eqnarray}
f &=& \sum_{j} \left[\sum_{i} (M^{-1})_{00} b_i N^{-1}_{i j} + \sum_{i} (M^{-1})_{01} N^{-1}_{i j}
\right](V_j - W_j^{SM}), \nonumber \\
  &=& \sum_{j} K(\theta_{j}) (V_j - W_j^{SM}).
\end{eqnarray} 
Computing~$K(\theta)$~to a cutoff radius of 1$\degr$ at 32 random points in the sky, we find that the profiles of $K(\theta)$ all share the same shape, as shown in Figure \ref{kernel}.
Hence we adopt the same polynomial form for the smoothing function~$S(\theta)$~to fit~$K(\theta)$, using the collective data from all the 32 sky points.
After filtering the residual map with $K(\theta)$, we minimize the noise of the map and as well obtain a better presentation of the point source flux distribution of the V - W sky. 

\begin{sidewaysfigure}
\label{mask}
\epsscale{0.95}
\plotone{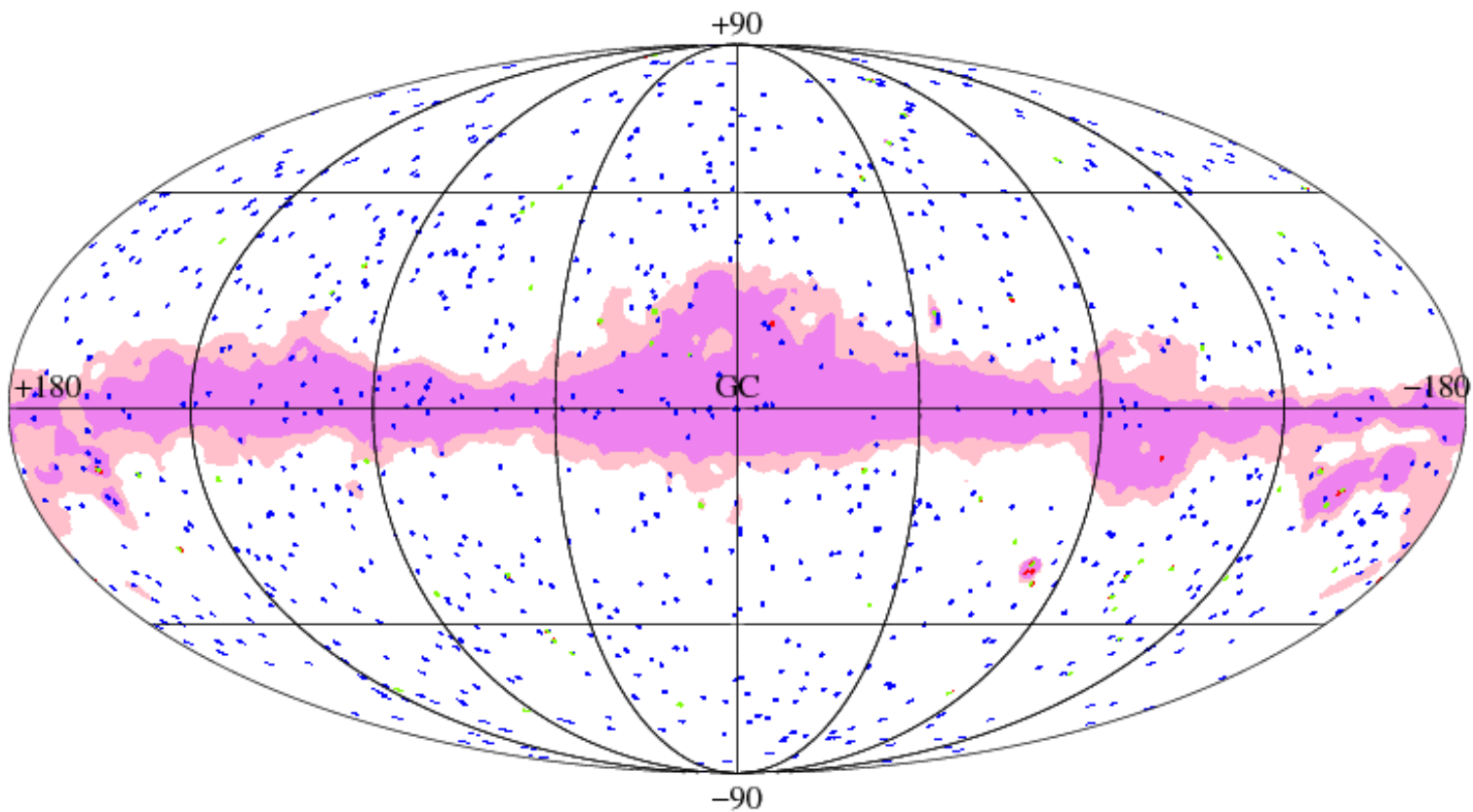}
\caption{An overview of the microwave sky.
The pink and violet shaded regions indicate the Kp0, Kp2 diffuse emission masks.
The blue dots give the point source mask.
The red and green dots together give the sources detected in our search; Green means it is included in the existed WMAP point source mask while red means not included.}
\end{sidewaysfigure}

We then look for peaks greater than $5\sigma$~in the $|b| > 10\degr$ region of the filtered map.
Note that the WMAP scan pattern gives non-uniform observation numbers among pixels:~greatest at the ecliptic poles, high at latitude $\pm45\degr$, and least in the ecliptic plane.
We thus divide the cut sky into 200 rings with different ecliptic latitudes and compute $\sigma$ separately within each ring.
When more than one $> 5\sigma$ detection lay within a 3-pixel by 3-pixel area, the brightest one is chosen as a source detection.
The position coordinates of a source are then carefully determined by fitting a point source profile in its 9-pixel neighborhood and locating the local maximum.
This process results in 31 point sources in the first-year maps and 64 point sources in the three-year co-added maps.
Since all the sources found in the first-year maps are also detected in the three-year maps, to avoid redundancy only the three-year point source catalog is given~(Table \ref{2band3yr}) and will hereafter be discussed in this paper.
Figure \ref{mask} offers an overview of the source distribution on the sky.
Some galactic sources are picked in our search due to the loose $|b| > 10\degr$ criterion.

\section{Identification}

We cross-correlate our source list with the latest WMAP point source catalog~(\citealt{2007ApJS..170..288H}).
If a source is within one half of the V-band beam width ($\theta^{V}_{HWHM} \sim 10^\prime$) to a cataloged WMAP source, we tag it.
We find 43 of our sources are present in the WMAP catalog, which leaves us 21 new detections.

\begin{figure}
\epsscale{0.62}
\plotone{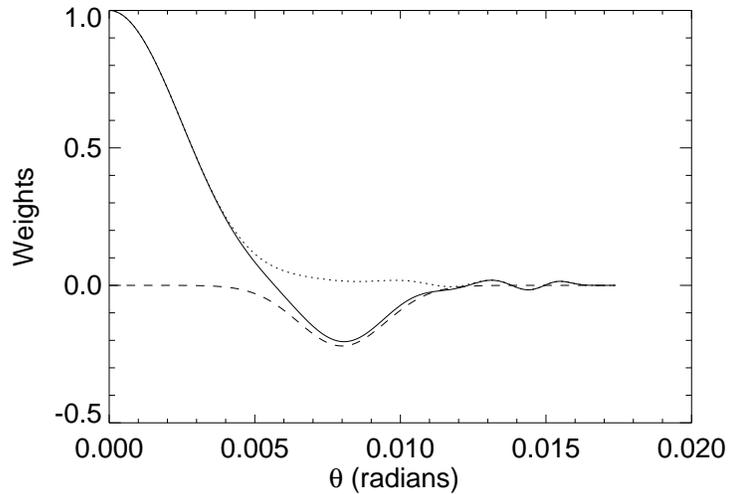}
\caption{The weighting function~(solid line) we introduced in our V-band flux estimator, which is constructed with a V-band beam matched central peak (dotted line) and a negative Gaussian (dashed line) outer ring.
Application of this weighting function provides a flux contrast enhancement of the point source from the background. 
\label{weight}}
\end{figure}

\begin{figure}
\epsscale{0.62}
\plotone{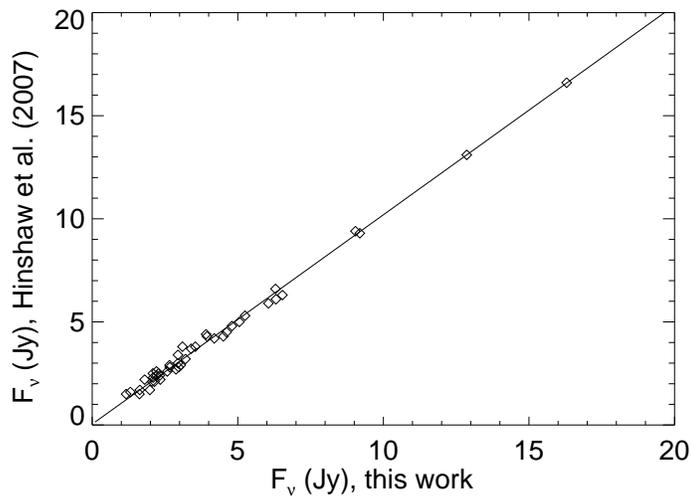}
\caption{Comparison of the point source V-band flux densities we initially computed in this work and those from the WMAP three-year release.
The diamonds are the 43 WMAP cataloged point sources we rediscovered in this work.
The linear fit has a slope of 1.02.
\label{compare}}
\end{figure}

The V-band flux densities of all our sources except for source 0540-0242 and 0734-4109 are calculated through
\begin{equation}
  F_{\nu} = \int I_{\nu}~\omega(\theta) \cos \theta d\Omega~=~\frac {\partial B_{\nu}}{\partial T} |_{T_{0}} \int T ~\omega(\theta) 
\cos \theta d\Omega,
\end{equation} 
where $T_{0}$ = 2.725$\pm$0.002~K is the CMB monopole temperature from COBE (\citealt{1999ApJ...512..511M});~$B_{\nu}$ is the Planck function; 
$T$ is the V-band temperature measurement (CMB dipole deducted);~The solid angle is integrated to a radius of 1 degree.
A weighting function $\omega(\theta)$~is introduced in the above equation, which is constructed with the V-band beam profile in the center and a negative Gaussian outer ring~(shown in Figure \ref{weight}).
The generated weights help to compensate the flux dilution caused by the big beam of WMAP and also to enhance the contrast of the point source flux to the background.
We enforce
\begin{equation}
   \int \omega(\theta) d\Omega = 0,
\end{equation}
so that any flux that spreads across the whole integration field will be ignored.
For source 0540-0242, since the bright source Orion B falls into the negative ring, using the standard weighting function would result in a negative flux.
The same concern holds for source 0734-4109, as it is located in the Gum Nebula complex and is surrounded by many radio emitting blobs.
Therefore, the V-band flux densities of these two sources are calculated using equation (13) without applying the weights.
We use the flux densities of the 43 WMAP cataloged sources (\citealt{2007ApJS..170..288H}) as a check of our flux estimates.
As seen in Figure \ref{compare}, these two sets of fluxes agree very well, with our values slightly off by a factor of 1.02$\pm$0.03 in the mean.
This is likely a consequence of the negative weights in our flux estimator.
We thus correct our flux estimates by scaling them up.
The uncertainty on the flux density is given by the statistical error of the fit and the rms of a simulated foreground-free CMB V-band map~(see~\S 4) under the same flux estimator.
We then search in the NASA/IPAC Extragalactic Database (NED\footnote{See \url{http://nedwww.ipac.caltech.edu/}.})~for radio sources within $\theta^{V}_{HWHM}$ to our sources and inspect their SEDs.
If the V-band flux of our source can be reasonably fitted into the SED of a known radio source, we tag it.
This procedure helps identify 15 more sources in our catalog and most of them are found to have a flat spectrum.

\begin{figure}
\epsscale{1}
\plottwo{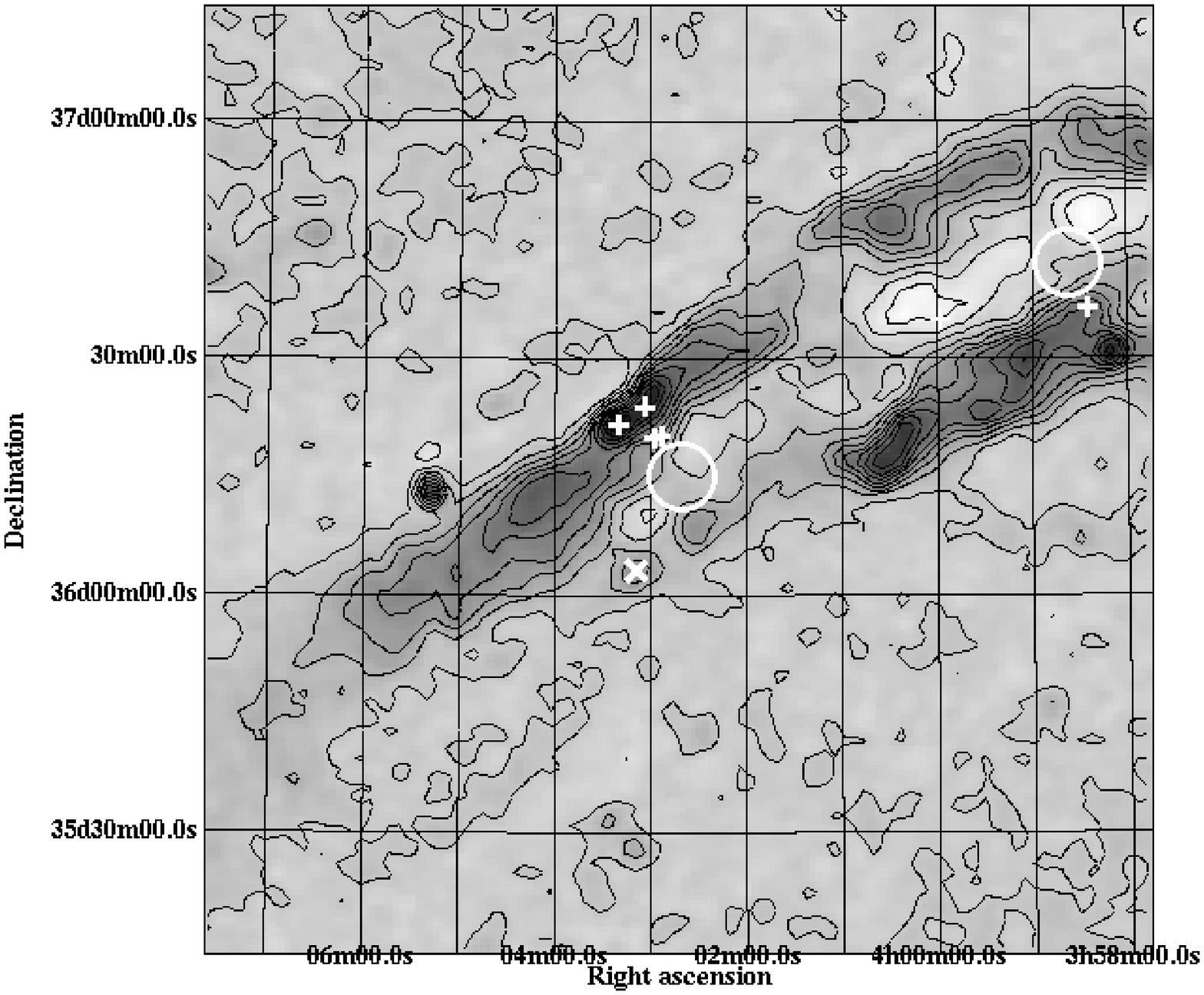}{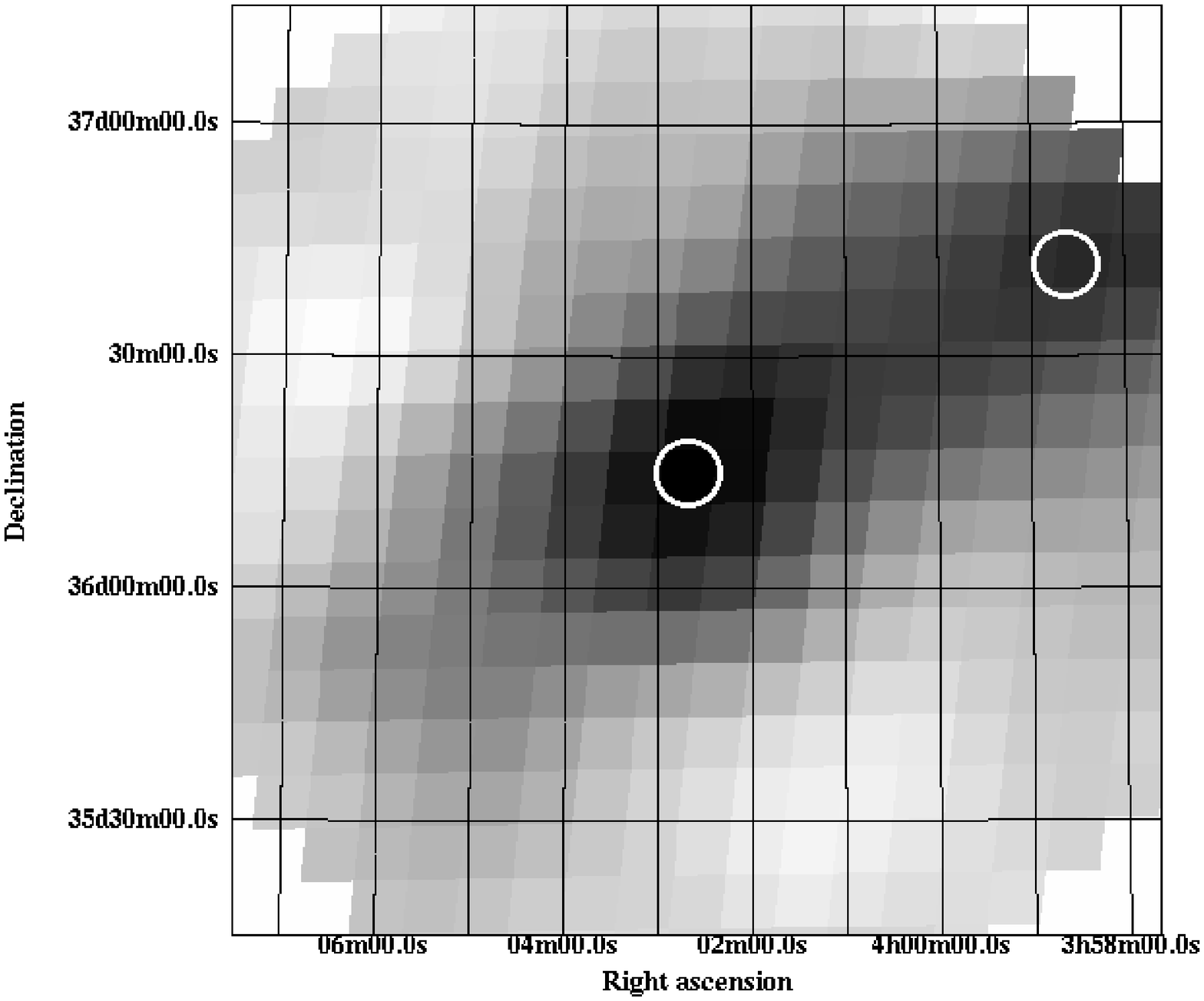}
\caption{\textbf{Left:}~$2\degr \times 2\degr$ GB6 map including our source 0402+3614 and 0358+3642.
White circle covers a 3$\sigma$ error area.
Cross points give MG2 and MG3 sources within one half the WMAP V-band beam width.
The X point is the closest GB6 source to our sources, which is 13.3$^\prime$ away from source 0402+3614.
\textbf{Right:}~Corresponding $2\degr \times 2\degr$ field in the filtered V-W map.}
\label{ngc1499}
\end{figure}

\begin{figure}
\epsscale{1}
\plottwo{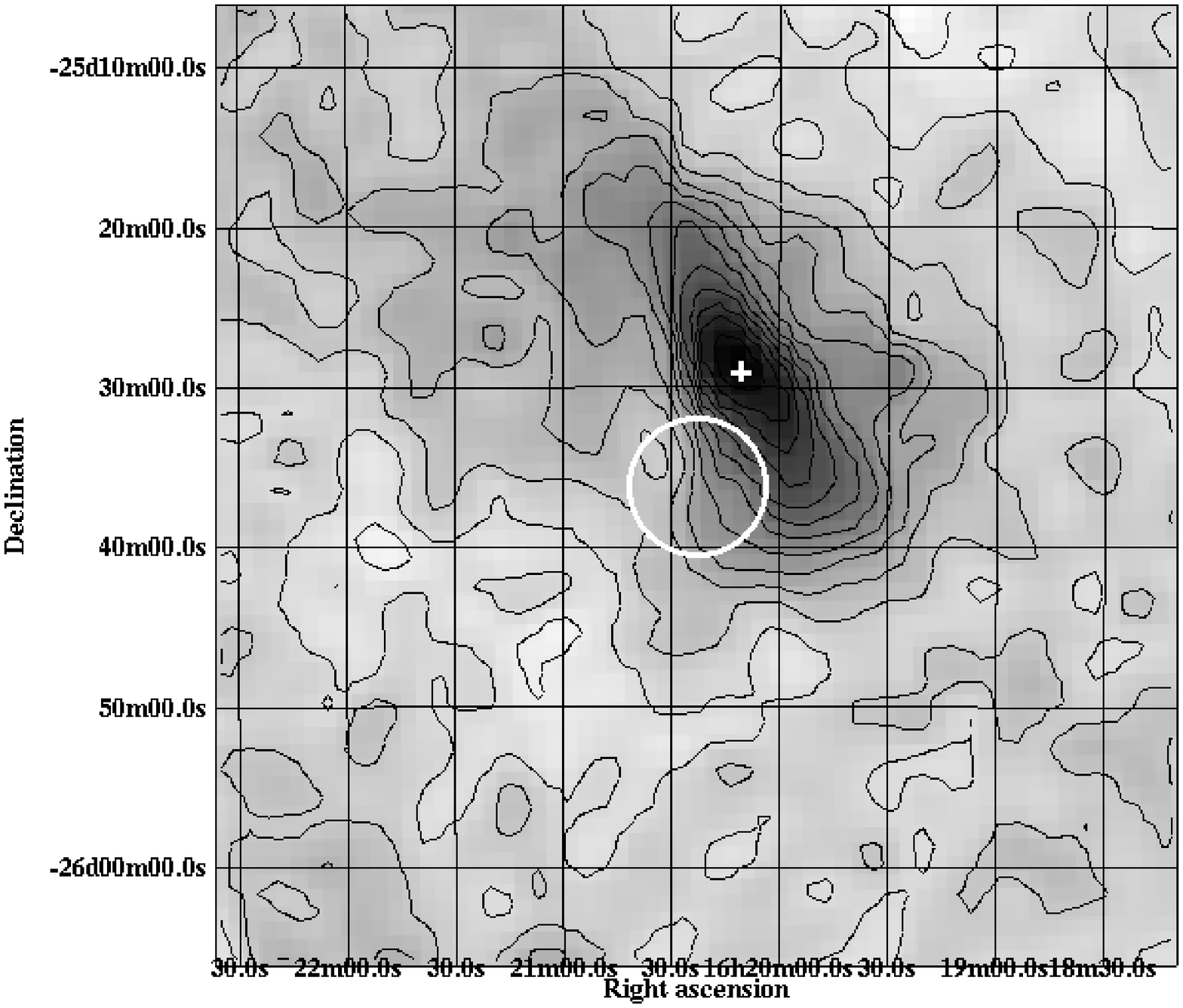}{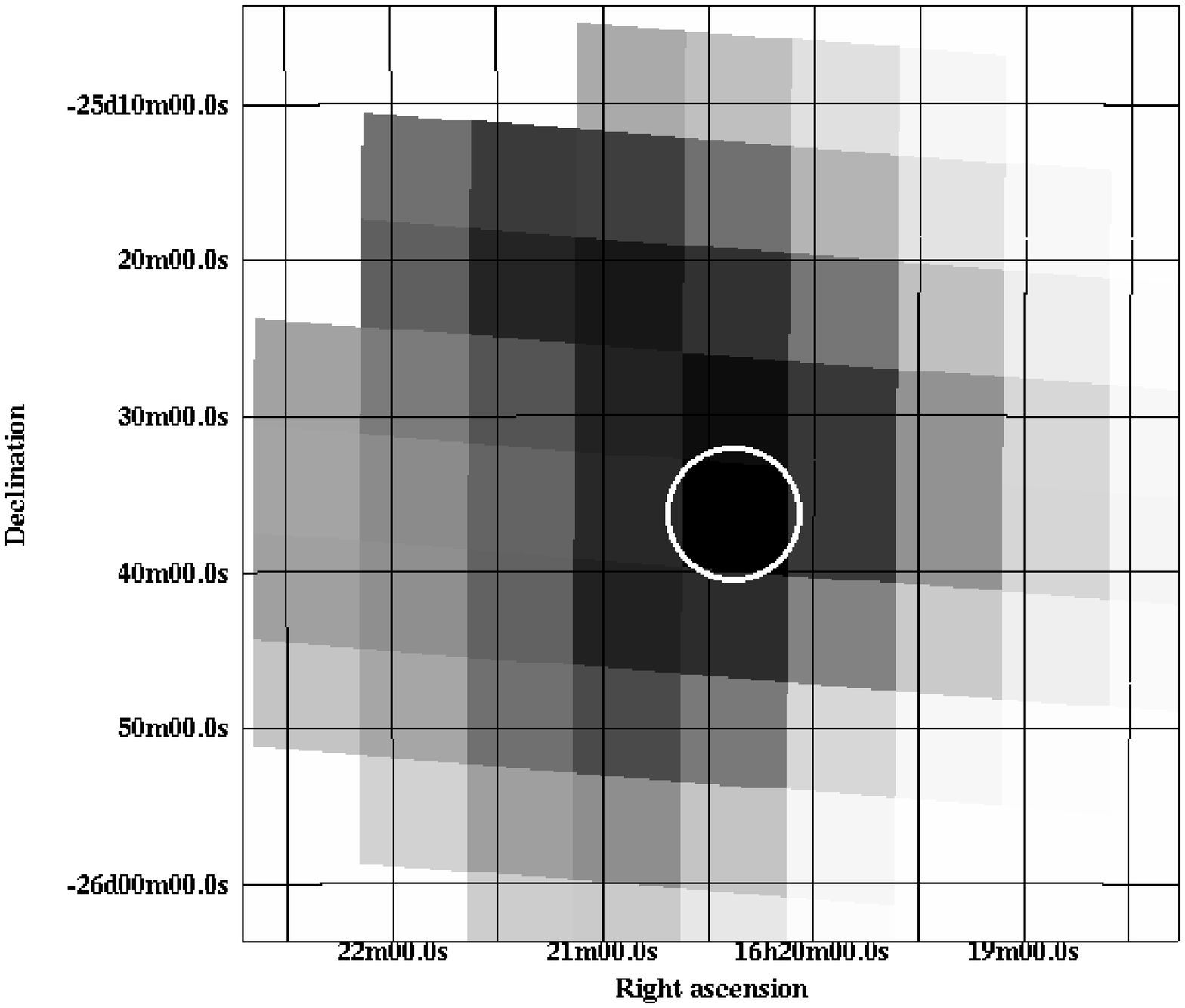}
\caption{\textbf{Left:}~$1\degr \times 1\degr$ PMN map centered on our source 1620-2536, marked by a 3$\sigma$ error circle.
Cross point indicates the PMN source within one half the WMAP V-band beam width.
\textbf{Right:}~Corresponding~$1\degr \times 1\degr$~field in the filtered V-W map.}
\label{rhooph}
\end{figure}

For source 0358+3642 and 0402+3614, several radio sources are found around our positions, but no one alone can explain the flux of either of our sources.
To have a direct view of this source-populated region, we obtain a $2\degr \times 2\degr$ GB6 map~(\citealt{1996ApJS..103..427G}) centered on the relatively brighter source 0402+3614, which immediately reveals that both sources lay within the boundary of the diffuse nebula NGC 1499~(Figure~\ref{ngc1499}).
We over-plot radio sources from 5 GHz surveys MG2~(\citealt{1990ApJS...72..621L}), MG3~(\citealt{1990ApJS...74..129G}) and GB6 on this radio image and find that although our sources are neither aligned with the strong radio emission bands nor coincident with any of these 5~GHz radio sources, they are largely within the WMAP V-band beam size.
We hence consider that our sources are caused by the beam smearing of WMAP and detected at the position where the composite flux reaches its maximum.
A double check in the corresponding WMAP V-W sky map (also shown in Figure~\ref{ngc1499})~confirms the general shape of the nebula and verifies that our sources are indeed the peaks in this blended emission area.
For source 1620-2536, although radio source PMN J1620-2529 is only 7.6$^\prime$ away, we cannot securely associate them as the only flux measurement of PMN J1620-2529 was made in 1994 (318 mJy at 4.85 GHz), and was one order of magnitude below the V-band flux of our source (2.2 Jy).
If they were the same source, a power spectrum index of $\sim$0.8 is required which indicates a negative value in the V-W map.
Moreover, this source failed to be detected in the latest Combined Radio All-sky Targeted Eight GHz Survey (CRATES, \citealt{2007ApJS..171...61H}).
However, we know from its coordinates that our source resides in the $\rho$~Ophiuchus nebula and a $1\degr \times 1\degr$ PMN map~(Figure~\ref{rhooph}, \citealt{1994ApJS...90..179G}) further shows that it coincides with a strong emission region within the nebula.
We double check the WMAP V-W sky map of the same field (also given in Figure~\ref{rhooph}) and find the brightness distribution is in good accordance with that in the PMN map.
Similarly, we find source 0734-4109 embedded in the Gum nebula complex, but no strong radio point source was reported in its immediate surroundings.
Therefore, no one-to-one connection can be made either.

\begin{figure}
\epsscale{0.65}
\plotone{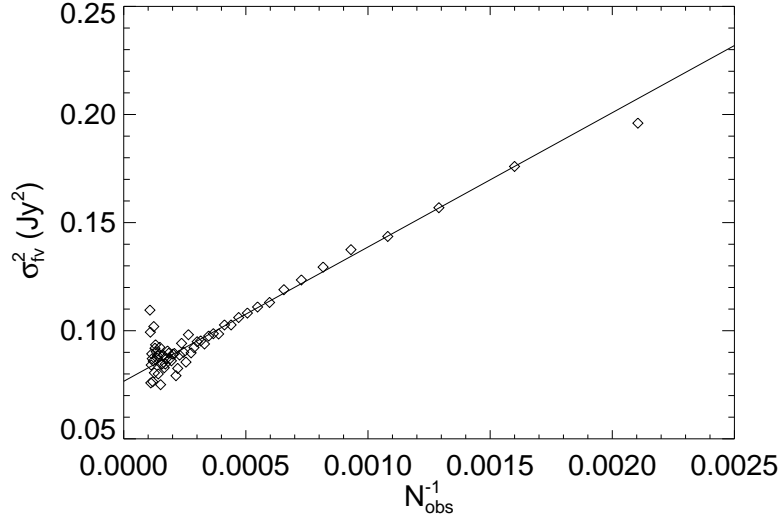}
\caption{The three years V-band flux variance dependence on observation numbers.
The diamonds are the median sample variances measured from each bin of map elements.
The solid line is a linear fit to the data. \label{variance}}
\end{figure}

Following the above steps, we solidly verify 58 sources in our catalog, and relate 4 more sources to known objects without attaching them to any single point source. For the two source candidates which we fail to associate with any known objects, snapshots were taken on source 0347+0733 at VLA 3.6~cm and CARMA 3~mm bands, and on source 1144-3902 at ATCA  3.5~cm and 6.25~cm bands during late 2006 to early 2007.
None of the above observations show any evidence of strong radio sources existing at the nominal positions.
This non-detection could be due to a large extent for these sources so that the small beams of the radio interferometers failed to catch them.
Or it could be due to the variability of these sources.
As we go through individual year WMAP data, we find that the V-band flux densities of these two sources did vary from year to year, given in Table \ref{fsyr}.
To judge the significance of this variability, we bin the map pixels (excluding those with $|b| \leqslant 10\degr$)~by different observation numbers, and fit the linear relation between median variance in each bin and the observation numbers (see Figure~\ref{variance})~to obtain the flux variance per observation~$\sigma^{2}_{0, median}$.
Since the three year flux variance can be approximated by a $\chi^{2}$ with 2 degrees of freedom, the mean variance per observation can then be calculated as $\sigma^2 = \langle \sigma_{0}^{2} \rangle/N_{obs}$, where~$\langle \sigma_{0}^{2} \rangle = \sigma^{2}_{0, median}/\ln2$.
We thus find that the flux variance of source~0347+0733 is 4.2 times as much as the expected level, and the variance for source 1144-3902 is 1.1 times the expected level.
In addition, as a 5$\sigma$ detection threshold can cause a few false detections in the 3 million pixels of WMAP, these two unidentified sources could also be spurious.

We find seven of our sources lying outside of the WMAP point source mask.
The 5 GHz counterparts of our sources are identified by cross-correlating our source catalog with the GB6~(\citealt{1996ApJS..103..427G}), PMN~(\citealt{1994ApJS...90..179G, 1995ApJS...97..347G}, \citealt{1994ApJS...91..111W, 1996ApJS..103..145W}), and \citet{1981A&AS...45..367K} catalogs.
We also check our source list with the results from two recent large sky radio survey: CRATES~(8.4 GHz, \citealt{2007ApJS..171...61H}) and the Australia Telescope 20 GHz Survey (AT20G, \citealt{2007arXiv0709.3485M}). If a source is cross-listed in the CRATES catalog or the Bright Source Sample (BSS) of AT20G survey, we mark it (see Table \ref{2band3yr}).

\section{Discussion}

\begin{figure}
\epsscale{0.65}
\plotone{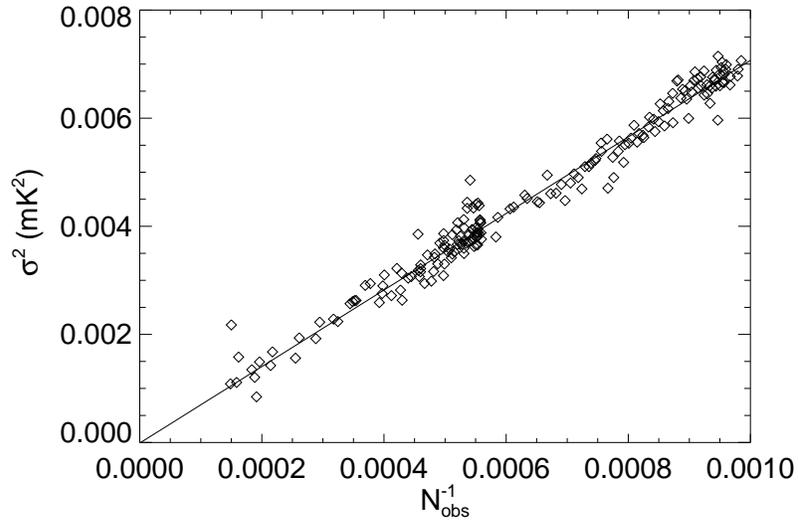}
\caption{Noise dependence on numbers of observation in the filtered V-W map.
The diamonds give the rms in each of the 200 rings at different ecliptic latitudes, and accordingly, different observation numbers.
Since all the convolutions we did to the maps are mathematically linear combinations, the linear relation between $\sigma$ and $N_{obs}^{-1/2}$ is preserved, and the solid line gives the fit.
\label{noise}}
\end{figure}

As specified in \S 2, we remove all the CMB signal from our search by subtracting the smoothed W-band map from the V-band map.
Hence, noise in the residual map comes primarily from the detector noise, which can be reduced as 1/$\sqrt{t}$ by integrating longer.
This trend is clearly demonstrated in Figure \ref{noise}, where we plot the noise dependence on the observation numbers in the filtered V-W map.
And we have also proved it by finding over $\sqrt{3}$ times as many sources in the three-year data as in the first-year data.
This is a distinct advantage of our method compared to the one used by the WMAP science team.
Recall that a main step in their source finding procedure is to filter the individual band maps with $b_{l}/(b_{l}^{2}C_{l}^{CMB}+C^{noise})$, where the CMB signal is a ``noise'' that does not integrate down by repeated observations.
To estimate the influence of this CMB ``noise", we generated monopole- and dipole-free CMB maps for each WMAP band using cosmological parameters favored by the combined data sets of WMAP 3 year data, 2dF Galaxy Redshift Survey~(\citealt{2005MNRAS.362..505C}), BOOMERanG and ACBAR~(\citealt{2006ApJ...647..813M}, \citealt{2004ApJ...600...32K}), CBI and VSA (\citealt{2004ApJ...609..498R}, \citealt{2004MNRAS.353..732D}), Sloan Digital Sky Survey~(\citealt{2004ApJ...606..702T}, \citealt{2005ApJ...633..560E}), Supernova Legacy Survey~(\citealt{2006A&A...447...31A}) and Supernova ``Gold Sample"~(\citealt{2004ApJ...607..665R}), assuming a $\Lambda$CDM model.
We then repeat the same procedure of weighting and filtering that \citet{2007ApJS..170..288H} did, and find the rms fluctuation level in the resultant maps to be 0.27 Jy, 0.41 Jy, 0.36 Jy, 0.27 Jy and 0.14 Jy for K-band through W-band, individually.
Considering that the detected sources are mostly about 1~Jy, this intrinsic CMB ``noise" may have caused some missing sources at a 5$\sigma$ level.

In addition, since the V- and W-band maps have comparatively higher angular resolution among the five WMAP bands, we achieve a higher positional accuracy in our search.
The 1$\sigma$ position uncertainty is 1.6$^\prime$ and 1.4$^\prime$ for the first-year and three-year result, respectively.
We also achieve a low detection threshold of 0.7~Jy at V-band in our search, attributed to the elimination of the CMB signal and the relatively low foreground contamination in the V- and W-band sky.
However, it is noticed that we did not detect all the sources with $F_V$ greater than 0.7~Jy in the WMAP three-year point source catalog.
This is because the above-mentioned subtraction step not only removed the CMB signal but also reduced the intensity of possible sources.
Assuming a flat spectrum, the W-band temperature of a source is about 1/2 the V-band temperature.
Therefore some strong sources in V-band could be diminished in the residual map. 

Besides the searches done by the WMAP science team, \citet{2007ChJAA...7..199N} and \citet{2007ApJS..170..108L}~independently carried out a point source search in the WMAP maps.
\citet{2007ChJAA...7..199N} applied the cross-correlation method developed by \citet{1983SSRv...36...61H} on the WMAP first year Q-band map and found 20 new sources at $|b| > 10\degr$, out of which we found 9.
In the non-blind search by~\citet{2007ApJS..170..108L}, they first generated an input catalog of 2491 sources from the 4.85~GHz GB6 and PMN surveys, 1.4~GHz NVSS survey~(\citealt{1998AJ....115.1693C}) and 0.843~GHz SUMSS survey~(\citealt{2003MNRAS.342.1117M}), and then applied the MHW2 wavelet filter~(\citealt{2006MNRAS.369.1603G}) to $14.6\degr \times 14.6\degr$ patches surrounding the sources and picked the $5\sigma$ detections.
15 new sources were found at V-band with $|b| > 10\degr$, out of which we also found 9.
These cross-detections, along with the 43 WMAP cataloged sources, prove that our approach is both competitive with and complementary to the other currently adopted methods for detecting point sources in the WMAP data.
Furthermore, our V-band flux estimates for these common sources are in general consistent with the estimates from other methods, which verifies the effectiveness of our flux estimator.

Combining our identified detections (excluding galactic sources) and the cataloged WMAP sources,
we modeled the V-band source count distribution $dN/dS$ as a power law 
$\kappa(S/[1\;\mbox{Jy}])^{-\beta}$, with~$\kappa = 17.6\pm3.3$ and $\beta = 2.4\pm0.1$. For Poisson-distributed sources, the contribution to the power spectrum by unresolved sources can be estimated by 
\begin{equation}
C = g(\nu)^2 \int_0^{S_c}~dS \frac {dN}{dS} S^2~\mu  {\rm K^2~sr},
\end{equation}
where $S_c$ is the flux limit and g($\nu$) = $(\partial B_{\nu}/ \partial T)^{-1}$.
Integrating this model with $S_c = 0.75\;\mbox{Jy}$ 
resulted in a contamination level of $C_V = (2.4\pm0.8) \times 10^{-3} \mu$K$^2$~sr for V-band.
Fitting this value to the power spectrum model of the unresolved sources, equation (1), we obtained a Q-band amplitude of $A = 0.012\pm0.004~\mu$K$^2$~sr, which is consistent with the estimates from both \citet{2007ApJS..170..288H} and \citet{2007arXiv0710.1873H}.

\section{Conclusions}
We present a CMB independent method for point source detection and apply it to the WMAP first-year and three-year maps.
The main result of this search is a point source catalog of 64 members, out of which 21 are newly detected in WMAP data.
Two sources remain unidentified.
We are not sure whether they are variable or extended sources.
More years WMAP data are needed to further identify them.
We demonstrate that our method is both competitive with and complementary to other currently adopted methods of point source searching in the WMAP data, with a low detection threshold as well as a good positional accuracy ($\sim 15\%$ of the W-band FWHM).
And our source list can be a good supplement to the existing WMAP point source catalog and the point source mask.
Contamination from unresolved point sources is at a negligible level of $(2.4\pm0.8) \times 10^{-3}~\mu$K$^2$~sr at WMAP V-band, which implies a correction value $A = 0.012\pm0.004~\mu$K$^2$~sr.

For most sources, with $F_{\nu} \propto \nu^{0}$, the K, Ka and Q bands have a higher signal to noise ratio and therefore can yield more source detections as proven in the search results from \citet{2007ApJS..170..288H}, \citet{2007ChJAA...7..199N} and \citet{2007ApJS..170..108L}.
However, part of the intention of this work was to look for interesting sources that might peak around the V and W bands.
Nevertheless, among all the identified sources in our catalog, we find 39 quasars, 10 galaxies, 9 HII regions and 4 other radio sources.
There is no sign of a significant different class of new objects.

We are planning to extend this work by including the WMAP Q-band data.
We anticipate having more source detections so that a more accurately model of the source count distribution and furthermore a better estimate of the contribution from unresolved point sources to the CMB power spectrum can be reached. 

\acknowledgments
We acknowledge the use of the Legacy Archive for Microwave Background Data Analysis (LAMBDA).
Support for LAMBDA is provided by the NASA Office of Space Science. We also want to acknowledge the use of the VLA, CARMA and ATCA.
We want to especially thank Philip Edwards for making the ATCA observation possible.
PMN and GB6 maps in this paper are retrieved from the SkyView Virtual Observatory, which is a service of the Astrophysics Science Division at NASA/  GSFC and the High Energy Astrophysics Division of the Smithsonian Astrophysical Observatory (SAO).

\clearpage

\begin{deluxetable}{lclccccl}
\tablecolumns{8}
\tabletypesize{\scriptsize}
\tablewidth{0pt}
\tablecaption{Point Source Catalog --Three Years \label{2band3yr}} 						
\tablehead{
\colhead{RA} & \colhead{DEC}  &
\colhead{WMAP ID} & \colhead{Type}  & \colhead{$F_V$} & \colhead{5GHz ID} & \colhead{Identified} & 
\colhead{Notes\tablenotemark{b}} \\
\colhead{hms} & \colhead{dms} & \colhead{} &
\colhead{} & \colhead{[Jy]} &  \colhead{} & \colhead{/Masked\tablenotemark{a}} & \colhead{} 
}
\startdata 
\hline
00 06 14 &  -06 23 44  & WMAP J0006-0622 & G   & 2.2$\pm$0.3 & PMN J0006-0623 & Y / Y & $\ddag$  \\
01 08 47 &   01 37 30  & WMAP J0108+0135 & QSO  & 2.4$\pm$0.3 & GB6 J0108+0135   &  Y / Y & $\ddag$ \\ 
01 36 58 &   47 51 27  & WMAP J0137+4752 & QSO  & 3.5$\pm$0.3 & GB6 J0136+4751   & Y / Y  & $\ddag$\\
02 10 46 &  -51 00 07  & WMAP J0210-5100  & QSO  & 2.7$\pm$0.3 & PMN J0210-5101  & Y / Y  & $\ddag,~\diamond$ \\
02 37 51 &   28 45 47  & WMAP J0237+2848  & QSO  & 2.7$\pm$0.3  & GB6 J0237+2848 & Y / Y  & $\ddag$ \\ 
03 19 52 &   41 30 35  & WMAP J0319+4131 &  G  & 4.9$\pm$0.4 & GB6 J0319+4130   & Y / Y  & $\ddag$,~Per A  \\
03 21 53 &  -37 05 50  & \nodata & RadioS  & 1.9$\pm$0.3 & PMN J0321-3711\tablenotemark{c}  &  Y / Y  & $\ast$,~For A west lobe \\
03 47 16 &   07 34 00  &   \nodata & \nodata & 0.7$\pm$0.3 &  \nodata &  N / N  & \nodata   \\
03 58 38 &   36 41 53  & \nodata  &  HII  & 2.4$\pm$0.3 &  \nodata  &   Y / Y  & In NGC 1499 \\ 
04 02 41 &   36 15 06  &  \nodata & HII  & 4.6$\pm$0.4 &  \nodata &  Y /  Y  &  In NGC 1499 \\ 
04 03 55 &  -36 05 25  & WMAP J0403-3604 & QSO  & 4.6$\pm$0.4 & PMN J0403-3605 & Y / Y &  $\ddag,~\diamond$ \\
04 23 20 &  -01 19 16  & WMAP J0423-0120  & QSO  & 9.2$\pm$0.5 & PMN J0423-0120 & Y / Y & $\ddag$  \\
04 40 39 &  -43 36 20  & WMAP J0440-4333 & QSO  & 2.4$\pm$0.3 & PMN J0440-4332 & Y / Y &  $\ddag,~\diamond$ \\   
04 55 45 &  -46 17 22  & WMAP J0455-4617 & QSO  & 3.6$\pm$0.3 & PMN J0455-4616 & Y / Y & $\ddag,~\diamond$ \\
04 57 15 &  -23 23 34  & WMAP J0457-2322 & QSO  &  1.9$\pm$0.3 & PMN J0457-2324 & Y / Y & $\ddag,~\diamond$ \\   
04 57 16 &  -66 25 31 &  \nodata  & HII  & 3.1$\pm$0.3  & PMN J0456-6624 & Y / Y  & $\dag$,~in LMC \\
05 17 10 &  -69 20 16  & \nodata & HII &  1.0$\pm$0.3 & PMN J0518-6914 & Y / N & $\dag$,~in LMC  \\
05 19 55 &  -45 48 59  &  WMAP J0519-4546  & G  & 3.1$\pm$0.3 & 1Jy 0518-45  & Y / Y & $\diamond$,~Pic A \\
05 22 18 &  -68 01 44  &  \nodata  &  HII &  1.7$\pm$0.3 & PMN J0522-6757 &  Y / N & $\dag,~\ddag$,~in LMC  \\ 
05 23 22 &  -36 26 25  &  WMAP J0523-3627 & G & 3.0$\pm$0.3 & PMN J0522-3628 & Y / Y & $\diamond$ \\   
05 35 16 &  -05 23 29  &   \nodata & HII & 290.8$\pm$8.0 & \nodata  & Y / Y  & Ori A \\
05 38 25 &  -69 07 37  &  \nodata & HII &  26.2$\pm$0.9 & PMN J0538-6905 &  Y / Y & $\dag,~\ddag$,~in LMC \\
05 38 35 &  -44 06 17  &  WMAP J0538-4405 & QSO  & 5.2$\pm$0.4 & PMN J0538-4405 & Y / Y & $\ddag,~\diamond$ \\
05 40 33 &  -02 42 15  & \nodata & RadioS  &  8.1$\pm$0.4 &  PMN J0541-0239\tablenotemark{c} &  Y / N  & $\ddag$ \\ 
05 41 43 &  -01 53 40  &  \nodata   & HII  & 45.6$\pm$1.4 & PMN J0541-0154 & Y / Y  & $\ddag$ \\
06 07 58 &  -06 25 52  &   \nodata  & HII  &  6.9$\pm$0.4 &  PMN J0607-0623  &  Y / Y  & $\dag,~\ast$ \\
06 09 42 &  -15 40 58  & WMAP J0609-1541 & QSO   & 2.2$\pm$0.3 & PMN J0609-1542 & Y / Y & $\ddag,~\diamond$ \\
06 35 17 &  -75 16 21  & WMAP J0635-7517 & QSO   & 3.0$\pm$0.3 & PMN J0635-7516 &  Y / Y & $\ddag,~\diamond$ \\ 
07 34 13 &  -41 09 31 & \nodata & RadioS  & 3.9$\pm$0.3 &  \nodata  &  Y / N  &  In Gum Nebula \\
08 36 25 &  -20 16 52  & WMAP J0836-2015 & QSO & 2.1$\pm$0.3 & PMN J0836-2017 & Y / Y & $\ddag,~\diamond$ \\
08 41 00 &   70 54 22  & WMAP J0841+7053 & QSO  & 1.7$\pm$0.3 & GB6 J0841+7053  & Y / Y & $\ddag$ \\
08 55 05 &   20 05 26  & WMAP J0854+2005  &  QSO  & 4.3$\pm$0.4 & 1Jy 0851+20 & Y / Y & $\ddag$ \\ 
09 09 09 &   01 26 01  & WMAP J0909+0118  &  QSO & 1.4$\pm$0.3 & GB6 J0909+0121  & Y / Y & $\ddag$ \\  
09 27 04 &   39 04 43  & WMAP J0927+3901  & QSO  & 4.0$\pm$0.3 &  GB6 J0927+3902 & Y / Y & $\ddag$ \\    
10 58 32 &   01 33 17  & WMAP J1058+0134  & QSO   & 4.0$\pm$0.3 & 1Jy 1055+01 & Y / Y & $\ddag$ \\
10 58 54 &  -80 03 13  &  WMAP J1059-8003 & QSO  & 2.4$\pm$0.3  & PMN J1058-8003 &  Y / Y &  $\ddag,~\diamond$ \\
11 44 20 &  -39 02 18  & \nodata & \nodata  & 0.8$\pm$0.3 &  \nodata  &  N / N  &  \nodata \\
12 29 09 &   02 02 59  & WMAP J1229+0203  & QSO  & 13.1$\pm$0.6 & GB6 J1229+0202 & Y / Y & $\ddag$ \\    
12 30 49 &   12 22 51  & WMAP J1230+1223 & G   & 9.4$\pm$0.5 & GB6 J1230+1223  &  Y / Y & Vir A \\
12 56 11 &  -05 47 31  & WMAP J1256-0547 & QSO  & 16.6$\pm$0.6 & PMN J1256-0547 & Y / Y &  $\ddag$ \\  
13 10 49 &   32 23 40  & WMAP J1310+3221 & QSO & 2.3$\pm$0.3 & GB6 J1310+3220  & Y / Y & $\ddag$ \\  
13 25 33 &  -42 59 19  &  \nodata  & G  &  25.7$\pm$0.9 &  PMN J1325-4257\tablenotemark{c}  & Y / Y & $\dag,~\ast,~\diamond$,~Cen A \\
13 37 43 &  -12 57 15  & WMAP J1337-1257 & QSO  & 6.4$\pm$0.4 & PMN J1337-1257 & Y / Y  & $\ddag$ \\ 
16 13 42 &   34 12 38  & WMAP J1613+3412  & QSO  & 2.7$\pm$0.3 & GB6 J1613+3412  & Y / Y &  $\ddag$ \\
16 20 23 &  -25 36 11  & \nodata & RadioS & 2.2$\pm$0.3 &  \nodata  &  Y / N  & In $\rho$~Oph nebula \\
16 35 03 &   38 07 21  & WMAP J1635+3807  & QSO & 4.7$\pm$0.4 & GB6 J1635+3808 & Y / Y & $\ddag$ \\
16 42 57 &   39 48 38  & WMAP J1642+3948  & QSO  & 5.4$\pm$0.4 & GB6 J1642+3948 & Y / Y  & $\ddag$ \\
17 03 31 &  -62 10 36  & WMAP J1703-6213  & G  & 2.1$\pm$0.3 & PMN J1703-6212 & Y / Y  &  $\ddag,~\diamond$ \\
17 20 30 &   -00 59 15 & \nodata  &  G  &  2.6$\pm$0.3 &  PMN J1720-0058   &  Y / Y &  $\dag,~\ast$ \\
17 33 00 &   -13 06 22 & \nodata & QSO & 4.2$\pm$0.4 &  PMN J1733-1304 & Y / Y & $\ast,~\ddag$ \\ 
17 43 57 &   -03 48 35 &  \nodata & QSO &  4.7$\pm$0.4  & PMN J1743-0350  & Y / Y & $\ast,~\ddag$ \\ 
17 51 41 &    09 41 10 & \nodata & QSO &  3.8$\pm$0.3 & GB6 J1751+0938  & Y / Y & $\dag,~\ast,~\ddag$ \\ 
17 53 43 &    28 47 04 & WMAP J1753+2848 & QSO  & 2.3$\pm$0.3 & GB6 J1753+2847 & Y / Y & $\ddag$ \\ 
18 06 38 &    69 50 48 & WMAP J1806+6949  & G & 1.2$\pm$0.3 & GB6 J1806+6949 & Y / Y & $\ddag$ \\ 
19 24 50 &   -29 14 38 &  \nodata & QSO  &  9.9$\pm$0.5 &  PMN J1924-2914  & Y / Y & $\dag,~\ast,~\ddag,~\diamond$ \\ 
19 27 39 &    73 57 42 & WMAP J1927+7357 &  QSO & 3.1$\pm$0.3 & GB6 J1927+7357 & Y / Y & $\ddag$ \\ 
21 48 10 &    06 57 04 & WMAP J2148+0657 & QSO  & 6.4$\pm$0.4 & GB6 J2148+0657 & Y / Y & $\ddag$ \\ 
21 57 11 &   -69 41 18 & WMAP J2157-6942 & G & 2.2$\pm$0.3 & PMN J2157-6941 & Y / Y & $\diamond$ \\ 
22 02 41 &    42 17 11 & WMAP J2202+4217 & QSO & 3.0$\pm$0.3  & GB6 J2202+4216 & Y / Y & $\ast$ \\
22 18 59 &   -03 33 18 & WMAP J2218-0335 & QSO & 1.7$\pm$0.3 & PMN J2218-0335    & Y / Y & $\ddag$ \\ 
22 25 41 &   -04 58 56 & WMAP J2225-0456 & QSO & 3.2$\pm$0.3  & PMN J2225-0457 & Y / Y & $\ddag$ \\ 
22 29 56 &   -08 33 29 & WMAP J2229-0832 & QSO & 3.3$\pm$0.3 & PMN J2229-0832 & Y / Y & $\ddag$ \\ 
22 53 57 &    16 09 07 & WMAP J2254+1608 & QSO & 6.7$\pm$0.4 & GB6 J2253+1608 & Y / Y & $\ddag$ \\ 
22 58 09 &   -27 59 58 & WMAP J2258-2757 & QSO & 6.2$\pm$0.4 & PMN J2258-2758 & Y / Y & $\ddag,~\diamond$ \\ 
\hline
\enddata
\tablenotetext{a}{Three-year WMAP point source mask is considered here.}
\tablenotetext{b}{$\dag$ and $\ast$ indicate the new point sources cross-detected in~\citet{2007ChJAA...7..199N} and~\citet{2007ApJS..170..108L}.~$\ddag$ and $\diamond$ indicate the source is included in the CRATES catalog~(\citealt{2007ApJS..171...61H})~and the AT20G BSS catalog~(\citealt{2007arXiv0709.3485M}),~respectively.}
\tablenotetext{c}{Indicates the source has multiple possible 5 GHz identifications. The brightest one is given here. }
\end{deluxetable}

\begin{deluxetable}{lcccccc}
\tablecolumns{7}
\tablewidth{0pt}
\tablecaption{Single Year V-band Fluxes of the Two Unidentified Sources \label{fsyr}} 						
\tablehead{
\colhead{RA} & \colhead{DEC}  & \colhead{$F_{V}^{yr1}$} & \colhead{$F_{V}^{yr2}$} & \colhead{$F_{V}^{yr3}$} & 
\colhead{$\sigma^{2}_{F}$} & \colhead{$\langle \sigma_{0}^{2} \rangle/N_{obs}$ } \\
\colhead{hms} & \colhead{dms} & \colhead{[Jy]} & \colhead{[Jy]} & \colhead{[Jy]} &  \colhead{[Jy$^{2}$]} & \colhead{[Jy$^{2}$]}
}
\startdata 
\hline
03 47 16 &   07 33 25  & 1.9$\pm$0.5 &  1.2$\pm$0.5  &  0.5$\pm$0.6 &  0.50  &  0.12  \\
11 44 20 &  -39 02 18  & 1.5$\pm$0.5 &  1.2$\pm$0.5  &  0.9$\pm$0.6 &  0.08  &  0.07  \\
\hline
\enddata
\end{deluxetable}

\end{document}